\documentclass[preprint,aps]{revtex4}
\usepackage{graphicx}
\usepackage{dcolumn}
\usepackage{bm}
\begin{document}
\preprint{APS-123/QED}

\title{ROTATIONAL DOPPLER EFFECT}
\author{Amit Halder}
\email{amithalder02@yahoo.co.in}
\affiliation{Department of Applied Sciences, Punjab Engineering College, Chandigarh-160012 (India)}
\date{\today}
\begin{abstract}
A monochromatic linear source of light is rotated with certain angular frequency and when such light is analysed after reflection then a change of frequency or wavelength may be observed depending on the location of the observer.  This change of frequency or wavelength is different from the classical Doppler effect \cite{doppler47} or relativistic Doppler effect \cite{einstein05}.  The reason behind this shift in wavelength is that a certain time interval observed by an observer in the rotating frame is different from that of a stationary observer.
\end{abstract}
\maketitle

A linear source of light (wavelength $\bm \lambda$ and frequency $\bm \nu$ ) is rotating at {\bf O} (Figure) and it sweeps an angle $\bm \Delta \theta$  in time $\bm \Delta t$, the angular frequency being $\bm \omega=\Delta \theta/\Delta t$, {\bf BQC} a reflector which reflects the light from {\bf O} to {\bf P}.  A ray of light ({\bf ray 1} ) traveling in the direction {\bf OB} reflected from the point {\bf B}, reaches the observer at {\bf P}.  The distance the light travels {\bf OBP} is {\bf r1}.  After a time $\bm \Delta t$ another ray {\bf OCP} ({\bf ray 2}) reaches the observer at {\bf P}, the distance it travels i.e. {\bf OCP} is {\bf r2}.  Let {\bf n } number of waves are emitted when the light sweeps through the angle $\bm \Delta \theta$  in time $\bm \Delta t$.  The observer at {\bf P} receives the same n-waves, the time difference between the first ray and the last ray (i.e. between {\bf ray 1} and {\bf ray 2}) is
\begin{figure}[h]
\vskip -0.25cm
\begin{center}
\resizebox{15.5cm}{6cm}{\includegraphics{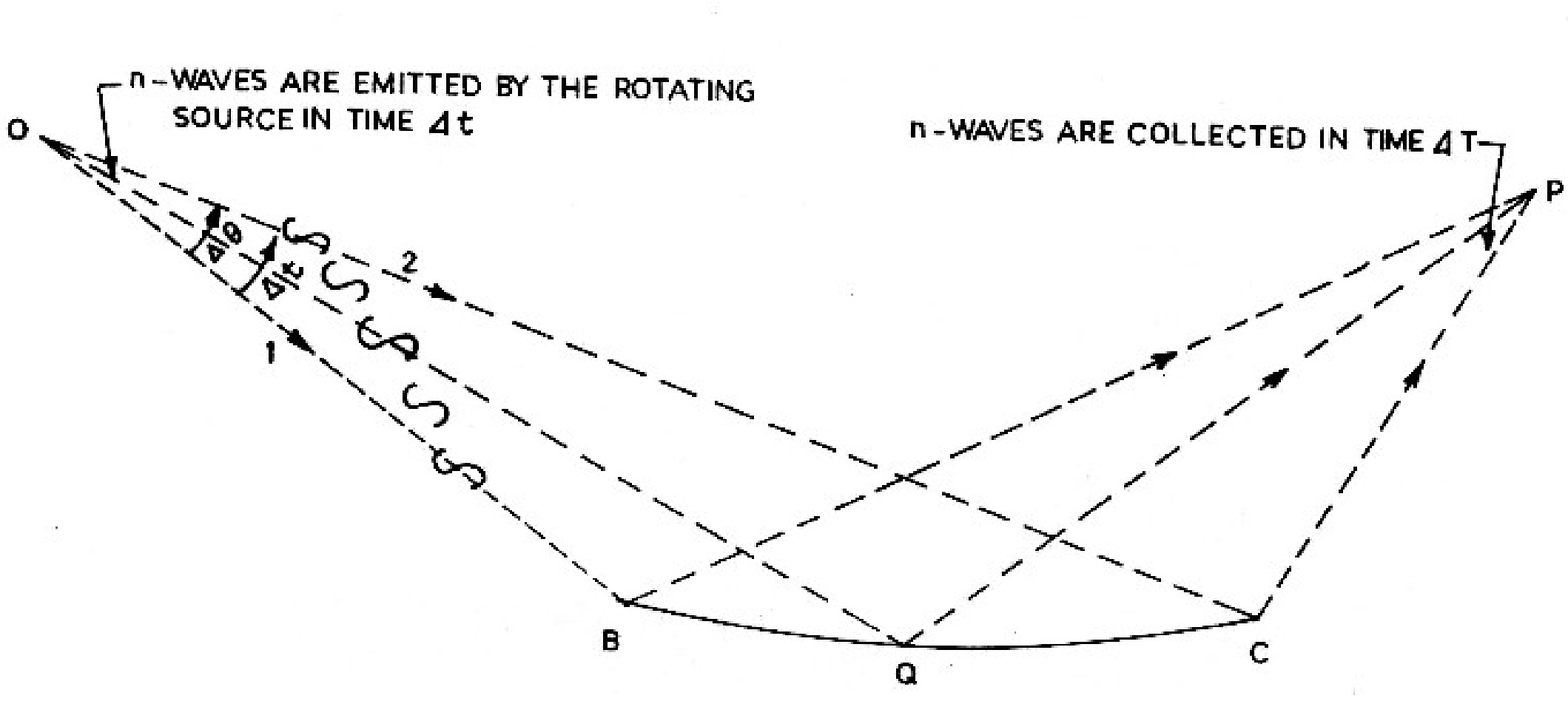}}
\end{center}
\label{fig}
\end{figure}
$$\bm  \Delta T  =  \Delta t  +  r2/c  -  r1/c  ~~~~ {\rm  (c~ is~ the~ velocity~ of~ light)} $$
	Because, {\bf ray 1} takes a time {\bf r1/c} and {\bf ray 2} takes a time $\bm \Delta t  +  r2/c$

$$\bm \Delta T  =  \Delta t  \pm   \frac{(r2  -  r1)}{C} $$
	+ ve sign when $\bm r2 > r1$ and  -ve sign when $\bm r2 < r1$
 
      $$\bm \Delta T  =  \Delta t   \pm    \delta r/c  {\rm~~ where~~} \Delta r  =  |r_2 - r_1|$$

It is to be noted that $\bm n$ waves are emitted by {\bf O} in time $\bm \delta t$    and the same $\bm n$ waves are collected or received by {\bf P} in time $\bm \Delta T$ and $\bm \Delta t \neq  \delta T$ because  $\bm r_2  \neq  r_1$.  In a special case $\bm r_2 =  r_1$ and $\bm \delta T  =  \delta t$.
	Since $\bm n$ waves are emitted in time $\bm \delta t$  therefore time period of one wave emitted by {\bf O} is $\bm \Delta t$ .  The frequency of the wave emitted by {\bf O}, $\bm \nu  =  n/ \Delta t$.

Similarly the same $\bm n$ waves are received at {\bf P} in time $\bm \Delta T$.  So time period of one wave received by {\bf P} will be $\bm \Delta T/n$.  The frequency of the received wave at {\bf P}, $\bm \nu'  = n/ \Delta T$
Since
$$\bm \Delta t   \neq \Delta T $$  
So 
$$\bm \nu  \neq \nu'$$ 
If
$$\bm  \lambda  =  c/\nu  {\rm ~ and~~}  \lambda'  =  c/\nu'$$ 
$$\bm   \lambda  =  c \Delta t /n$$  
and
$$\bm \lambda'  =   c \Delta T/n$$
Changes in wavelength, 
$$\bm  \Delta \lambda  =  \lambda'  -  \lambda  =  c/n  ( \Delta T  -  \Delta t )$$
$$\bm   \Delta \lambda  =  c/n  ( \pm  \delta  r/c )$$
$$\bm   \Delta \lambda  =  \pm  \delta  r/n $$ 
$$\bm	\Delta \lambda/\lambda    =   \pm \Delta r/c \Delta t$$
and
$$\bm  \Delta \lambda / \lambda'    =   \pm \Delta r/c \Delta T$$
           \centerline{ \bf  + ve sign means an increase in wavelength} \\
           \centerline{\bf  - ve sign means a decrease in wavelength.}\\
The change in wavelength depends on $\bm \Delta  r$  i.e. the difference between {\bf r1} and {\bf r2}.  So, the change in wavelength depends on the location of the point {\bf P}.  Also there is no shift in wavelength when  $\bm r2  = r 1$.\\
The above expression for shift in wavelength is different from the classical expression for Doppler effect as well as from the relativistic Doppler effect (Transverse Doppler effect).

It is evident in the present circumstance that the time interval recorded by a rotating frame at {\bf O} is $\bm \Delta t$ whereas the same time interval is recorded by a stationary frame is $\bm \Delta T$, they are related by the equation :  $\bm \Delta T  =  \Delta t   \pm   \Delta  r/c$.  again $\bm \Delta T$   depends on $\bm \Delta r$ i.e. location of the point {\bf P}.


\begin{thebibliography}{99}
\bibitem{doppler47} Christan Doppler, Abh. K"ngl .  b"hm. Geselsch,  2 (1842), 446 and Pogg. Ann. 68  (1847). 1
\bibitem{einstein05} A. Einstein, ``On the electrodynamics of moving bodies'',  Ann. Physik, 17 (1905) 891.
\end{thebibliography}

\end{document}